\documentclass[lettersize,journal]{IEEEtran}
\usepackage{amsmath,amsfonts}
\usepackage{algorithmic}
\usepackage{algorithm}
\usepackage{array}
\usepackage[caption=false,font=normalsize,labelfont=sf,textfont=sf]{subfig}
\usepackage{textcomp}
\usepackage{stfloats}
\usepackage{url}
\usepackage{verbatim}
\usepackage{graphicx}
\usepackage{cite}
\usepackage{amssymb}
\usepackage{xcolor}
\usepackage{color}
\usepackage{titlesec}
\usepackage{graphicx} 
\usepackage{epsfig}
\usepackage{epstopdf} 
\titlespacing{\subsection}{0pt}{*0.8}{*0.8} 
\titlespacing{\section}{0pt}{*0.8}{*0.9} 
\DeclareMathOperator*{\argmax}{arg\,max}
\DeclareMathOperator*{\argmin}{arg\,min}

\begin{document}

\title{Blind Channel Estimation for RIS-Assisted Millimeter Wave Communication Systems}

\author{Dianhao Jia, Wenqian Shen,~\IEEEmembership{Member,~IEEE}, Jianping An,~\IEEEmembership{Senior Member,~IEEE}, and Byonghyo Shim, ~\IEEEmembership{Senior Member,~IEEE}
\thanks{Copyright (c) 2025 IEEE. Personal use of this material is permitted. However, permission to use this material for any other purposes must be obtained from the IEEE by sending a request to pubs-permissions@ieee.org. This work was supported in part by the Natural Science Foundation of Beijing, China under Grant 4252010, in part by the Fundation by National Key Laboratory of Science and Technology on Space Microwave under Grant HTKJ2024KL504006 and in part by the NRF grant funded by the Korea government (MSIT-RS-2022-NR070834). \textit{(Corresponding author: Wenqian Shen.)}}
\thanks{Dianhao Jia is with the School of Cyberspace Science and Technology, Beijing Institute of Technology, Beijing 100081, China (e-mail: 3220231775@bit.edu.cn). }
\thanks{Wenqian Shen and Jianping An are with the School of Information and Electronics, Beijing Institute of
Technology, Beijing 100081, China (e-mail: shenwq@bit.edu.cn; an@bit.edu.cn). }
\thanks{Byonghyo Shim is with the Institute of New Media and Communications and Department of Electrical and Computer Engineering, Seoul National University, Gwanak-gu 151-742, South Korea (e-mail: bshim@snu.ac.kr).}}

\maketitle

\begin{abstract}
In the research of RIS-assisted communication systems, channel estimation is a problem of vital importance for further performance optimization. In order to reduce the pilot overhead to the greatest extent, blind channel estimation methods are required, which can estimate the channel and the transmit signals at the same time without transmitting pilot sequence. Different from existing researches in traditional MIMO systems, the RIS-assisted two-hop channel brings new challenges to the blind channel estimation design. Hence, a novel blind channel estimation method based on compressed sensing for RIS-assisted multiuser millimeter wave communication systems is proposed for the first time in this paper. Specifically, for accurately estimating the RIS-assisted two-hop channel without transmitting pilots, we propose a block-wise transmission scheme. Among different blocks of data transmission, RIS elements are reconfigured for better estimating the cascade channel. Inside each block, data for each user are mapped to a codeword for realizing the transmit signal recovery and equivalent channel estimation simultaneously. Simulation results demonstrate that our method can achieve a considerable accuracy of channel estimation and transmit signal recovery.
\end{abstract}

\section{Introduction}
\IEEEPARstart{D}{ue} to its capability to enhance the power of received signals by reconfiguring the amplitude and phase of the reflected signals, reconfigurable intelligent surface (RIS) has received much attention as a candidate  technology for 6G recently. In order to enjoy the full potential of an RIS, accurate channel statement information (CSI) is required for solving various kinds of optimization problems \cite{ref1}. However, elements of RIS are usually installed at a large scale which can cause an unaffordable cost for estimating the channel. Over the years, many works on channel estimation for RIS-assisted systems have already been carried out. In \cite{ref2}, a two-timescale channel estimation framework dividing the cascaded channel into a quasi-static base station (BS)-RIS channel and a low-dimension RIS-user equipment (UE) channel has been proposed. In \cite{ref3}, by utilizing a sensing RIS that integrates power detectors, a dimension-independent CSI acquisition scheme is proposed to significantly reduce the channel estimation overhead. In \cite{ref4}, the cascaded channel estimation problem is converted into a sparse signal recovery problem using the inherent sparsity in mmWave channels, and compressed sensing (CS) method has been applied to solve the problem. 
These approaches reduce the pilot overhead by using the channel inherent characteristics. 

In the traditional communication systems without RIS, plenty of researches have been carried out on blind channel estimations where statistical characteristics of received signal are exploited to estimate the wireless channel without pilot sequence. In \cite{EVD_BCE}, an eigenvalue decomposition-based blind channel estimation has been proposed by exploiting the asymptotic orthogonality of channel vectors in MIMO systems. Furthermore, as the developing requirement of Internet-of-Thing and machine-type communications, a new paradigm of communication named Unsourced Multiple Access (UMA) has received attention\cite{UMA_1}. Due to the massive number of devices and the uncoordinated characteristic, blind receiver plays an important role in UMA systems while the conventional training based methods are not feasible\cite{BSD_mMIMO}\cite{UMA_BCE}. 
In massive MIMO system, \cite{BSD_mMIMO} performs channel estimation and symbol detection simultaneously by factorizing the received signal by utilizing the channel sparsity. In \cite{UMA_BCE}, the transmit signal is generated from a codebook and the codeword sparsity as well as channel sparsity are exploited to propose a blind receiver. However, conventional blind estimation methods are not applicable to RIS-assisted systems because of the cascade structure.
 
Moreover, in the RIS-assisted systems, several semi-blind channel estimation techniques have been proposed. In \cite{ref5}, third-order tensor-based model has been proposed along with the receiver algorithm. By utilizing the PARATUCK decomposition, the proposed algorithm estimates two involved channel matrices and the symbol matrix alternatively. In \cite{SBCE_RIS_EM}, a semi-blind iterative expectation maximization (EM)-based algorithm has been proposed to estimate the cascaded RIS-assisted channel by assuming Gaussian priori on the transmit symbols. Although the length of pilot can be reduced considerable, a certain length of pilot is still required to eliminate the scaling ambiguities. 

In this paper, inspired by \cite{UMA_BCE}, we propose the blind channel estimation method for RIS-assisted multiuser mmWave systems. In order to facilitate the blind estimation of RIS-assisted cascaded channel, we propose the block-wise transmission scheme where the transmit signal in each block is generated from a common codebook. As the phase shift matrix of RIS is adjustable, we divide the channel estimation time interval into multiple blocks while RIS elements are reconfigured among blocks, which makes the channel environment different among blocks. Then the cascaded two-hop channel as well as the transmit signal are recovered by CS-based algorithms. We also provide the analysis of complexity and overhead of our proposed method comparing with traditional pilot-based as well as semi-blind channel estimation methods. Simulation results shows that the proposed blind channel estimation method achieves a considerable performance with low overhead. 
\section{System Model}
\subsection{Transmit Signals}
In order to encode the transmit signals for users, we generate a common codebook $\mathbf{C}\in \mathbb{C}^{M\times N}$ with each element following Gaussian distribution, i.e., $\mathbf{C}\sim \mathcal{CN}(0,\mathbf{1})$, for all the $K$ users where $M$ is the length of each codeword and $N$ is the total number of codewords in this codebook. A code bit vector to be transmitted by the $k$-th user can be expressed as $\mathbf{b}_k\in \{0,1\}^{M_b}$ and the size of the codebook satisfies $2^{M_b}=N$. Thus, vector $\mathbf{b}_k$ is able to be mapped to a codeword $\mathbf{x}_k \in \mathbb{C}^{M\times 1}$, which is the $n_k$-th column in codebook $\mathbf{C}$, where $n_k$ is the decimal integer with a binary expansion given by $\mathbf{b}_k$. Hence, all the possible code bit vectors $\mathbf{b}_k$ are one-to-one mapped to those $N$ codewords in $\mathbf{C}$. So that we can use an indicator vector $\boldsymbol{\gamma}_k \in \mathbb{C}^{N\times 1}$ to describe the mapping process and represent the transmit signal of user-$k$ as 
\begin{equation}\label{x=Cr}
  \mathbf{x}_k=\mathbf{C}\boldsymbol{\gamma}_k
\end{equation}
where the $n_k$-th element of $\boldsymbol{\gamma}_k$ is set to one and the rest are all zeros. Hence, transmit signal $\mathbf{x}_k$ is determined through a sparse vector $\boldsymbol{\gamma}_k$.
\subsection{Channel Model}
We consider the uplink of a RIS-assisted multiuser mmWave system, where a RIS reflects signals from $K$ single-antenna users to the BS equipped with $N_{\text{B}}$ antennas. We assume that the RIS is composed of $N_{\text{R}}$ reflecting elements, each of them can adjust the amplitude and the phase of the incident signal separately. Let $\psi_i=\beta_i e^{j\theta_i}(i=1,\ldots ,N_{\text{R}})$ be the reflecting coefficient related to the $i$-th element of the RIS, where $\beta_i\in[0,1]$ and $\theta_i\in(0,2\pi]$ denote the amplitude reflection and phase shift coefficients, respectively. For simplicity, we assume $\beta_i=1$ in this paper\cite{ref4}.

According to the Saleh-Valenzuela model, narrow band mmWave channel models of the RIS-BS channel $\mathbf{H}_{\text{RB}}\in\mathbb{C}^{N_{\text{R}}\times N_{\text{B}}}$, $k$-th UE-RIS channel $\mathbf{h}_{\text{RU},k}\in\mathbb{C}^{N_{\text{R}}\times 1}$ can be expressed as \cite{ref7}
\begin{equation}
\label{H_RB} 
\vspace{-0.5em}
\mathbf{H}_{\text{RB}}=\sqrt{\frac{N_{\text{R}} N_{\text{B}}}{L_{\text{RB}}}}\sum_{l_1=1}^{L_{\text{RB}}} \rho_{l_1} \mathbf{a}_{\text{R}}(\phi_{l_1})\mathbf{a}_{\text{B}}^H(\varphi_{l_1})
 \end{equation}
\begin{equation}\label{H_RUk}
\vspace{-0.5em}
   \mathbf{h}_{\text{RU},k}=\sqrt{\frac{N_{\text{R}}}{L_{\text{RU},k}}}\sum_{l_{2,k}=1}^{L_{\text{RU},k}} \varrho_{l_{2,k}} \mathbf{a}_{\text{R}}(\varphi_{l_{2,k}})
\end{equation}
where $L_{\text{RB}}$ is the number of propagation paths of RIS-BS channel and $L_{\text{RU},k}$ is the number of propagation paths of the channel from the $k$-th user to RIS. Variable $\rho_{l_1}$ and $\varrho_{l_2,k}$ represent the complex path gains. Suppose that uniform linear arrays (ULAs) are equipped at both BS and RIS, the array response vectors at RIS can be expressed as $\mathbf{a}_{\text{R}}(\phi_{l_1})=\frac{1}{\sqrt{N_{\text{R}}}}\left[1\:e^{j2\pi d\cos(\phi_{l_1})/\lambda}\:\ldots\:e^{j2\pi (N_{\text{R}}-1)d\cos(\phi_{l_1})/\lambda}\right]^T$, where $\phi_{l_1}$ denotes the angle of departure (AoD) at RIS, $\lambda$ denotes the wavelength of the signal and $d=\lambda/2$ is the inter-element spacing of antennas. The array response vector $\mathbf{a}_{\text{B}}(\varphi_{l_1})$ and $\mathbf{a}_{\text{R}}(\varphi_{l_{2,k}})$ can be similarly defined, with $\varphi_{l_1}$ and $\varphi_{l_{2,k}}$ being the angle of arrival (AoA) at BS and RIS, respectively.

Due to the scattering property of mmWave channels, the mentioned two channels can be approximately rewritten with virtual channel representations as \cite{ref4}
\vspace{-0.2cm}
\begin{equation}\label{SparseFormH_RB}
  \mathbf{H}_{\text{RB}}=\mathbf{F}_{\text{RB}}\mathbf{D}_{\text{RB}}\mathbf{F}_{\text{B}}^H
\end{equation}
\vspace{-10pt}
\begin{equation}\label{SparseFormH_RUk}
  \mathbf{h}_{\text{RU},k}=\mathbf{F}_{\text{RU}}\mathbf{d}_{\text{RU},k}
  \vspace{-0.1cm}
\end{equation}
\vspace{-0.1cm}
where $\mathbf{F}_{\text{B}}=[\mathbf{a}_{\text{B}}(\varphi_1),\ldots,\mathbf{a}_{\text{B}}(\varphi_{G_\text{B}})]\in \mathbb{C}^{N_{\text{B}}\times{G_{\text{B}}}}$ is the array response matrix at BS which is composed of $G_{\text{B}}$ array response vectors, and $\mathbf{F}_{\text{RB}}=[\mathbf{a}_{\text{R}}(\phi_1),\ldots,\mathbf{a}_{\text{R}}(\phi_{G_{\text{RB}}})]\in \mathbb{C}^{N_{\text{R}}\times{G_{\text{{RB}}}}}$, $\mathbf{F}_{\text{{RU}}}=[\mathbf{a}_{\text{R}}(\phi_1),\ldots,\mathbf{a}_{\text{R}}(\phi_{G_{\text{{RU}}}})]\in \mathbb{C}^{N_{\text{R}}\times{G_{\text{{RU}}}}}$ are the array response matrices at RIS in the RIS-BS channel and the UE-RIS channel, respectively. For simplicity, we assume that the numbers of discrete grids at RIS have $G_{\text{RB}}=G_{\text{RU}}=G_{\text{R}}$. $\mathbf{D}_{\text{RB}}\in \mathbb{C}^{G_{\text{R}}\times{G_{\text{B}}}}$ and $\mathbf{d}_{\text{RU},k}\in \mathbb{C}^{G_{\text{R}}\times{1}}$ are the equivalent sparse channel in angular domain with $L_{\text{RB}}$ and $L_{\text{RU},k}$ non-zero entries corresponding to the channel path gains, respectively.
\subsection{Received Signals}
 Assuming that the direct links from users to BS are blocked by obstacles in the wireless environment, the signal $\mathbf{Y}\in\mathbb{C}^{M\times N_{\text{B}}}$ received by the BS during time $M$ is given by
 \vspace{-0.2cm}
\begin{eqnarray}\label{y(m)}
\small
  \mathbf{Y}
  &=&\sum_{k=1}^{K}\mathbf{Y}_k
  =\sum_{k=1}^{K}\mathbf{x}_k\mathbf{h}_{\text{RU},k}^H\text{diag}
  (\boldsymbol{\psi})\mathbf{H}_{\text{RB}}+\mathbf{N}\nonumber\\
   &=&[\mathbf{x}_1,\ldots,\mathbf{x}_K]   \left[  \begin{array}{cc}
    \boldsymbol{\psi}^{T} \mathbf{H}_1 \\
   \vdots \\ 
     \boldsymbol{\psi}^{T} \mathbf{H}_K\end{array}\right]+\mathbf{N}
\end{eqnarray}
where $\mathbf{x}_k$ denotes the transmit signal from user-$k$, $\mathbf{N}\in \mathbb{C}^{M\times N_{\text{B}}}$ is the additive white Gaussian noise with zero mean and variance $\sigma^2$, and $\boldsymbol{\psi}=[\psi_1,\ldots,\psi_{N_\text{R}}]^T\in \mathbb{C}^{N_\text{R}\times 1}$ denotes the phase shift vector of RIS, $\mathbf{H}_k=\text{diag}(\mathbf{h}_{\text{RU},k}^{*})\mathbf{H}_{\text{RB}}\in \mathbb{C}^{N_{\text{R}} \times N_{\text{B}}}$ denotes the cascaded channel from the $k$-th user to the BS via RIS, where the notation $(\cdot)^*$ denotes the conjugate of a vector or matrix. Combined with (\ref{SparseFormH_RB}) and (\ref{SparseFormH_RUk}), the cascaded channel can be written into a sparse form as
\begin{eqnarray}\label{H_kSparse}
  \mathbf{H}_k&=&\text{diag}(\mathbf{h}_{\text{RU},k}^{*})\mathbf{H}_{\text{RB}} \nonumber\\
  &\overset{(a)}{=}&(\mathbf{F}_{\text{RU}}^*\bullet\mathbf{F}_{\text{RB}})(\mathbf{d}^{*}_{\text{RU},k}\otimes
  \mathbf{D}_{\text{RB}})\mathbf{F}_{\text{B}}^H \nonumber\\
  &=&\widetilde{\mathbf{F}}_{\text{R}}\widetilde{\mathbf{D}}_k\mathbf{F}_{\text{B}}^H \nonumber \\
  &\overset{(b)}{=}&\mathbf{F}_{\text{R}}\mathbf{D}_k\mathbf{F}_{\text{B}}^H
\end{eqnarray}
where $\bullet$ denotes the “transposed Khatri-Rao product”, and $\otimes$ denotes the Kronecker product. The equation $(a)$ follows from the Kronecker product property \cite{Matrix} and we defined $\widetilde{\mathbf{F}}_{\text{R}}\triangleq\mathbf{F}_{\text{RU}}^{*}\bullet\mathbf{F}_{\text{RB}} \in \mathbb{C}^{N_{\text{R}} \times G_{\text{R}}^2}$ and $\widetilde{\mathbf{D}}_k\triangleq\mathbf{d}^{*}_{\text{RU},k}\otimes\mathbf{D}_{\text{RB}}\in \mathbb{C}^{G_{\text{R}}^2 \times G_{\text{B}}}$. In equation $(b)$, based on the column repetitiveness of $\widetilde{\mathbf{F}}_{\text{R}}$, we can represent $\mathbf{F}_{\text{R}}\in \mathbb{C}^{N_\text{R}\times G_\text{R}}$ as the first $N_\text{R}$ columns of $\widetilde{\mathbf{F}}_{\text{R}}$ and $\mathbf{D}_k \in \mathbb{C}^{G_{\text{R}} \times G_{\text{B}}}$ as a merged version of $\widetilde{\mathbf{D}}_k$ \cite[\textit{Proposition 1}]{ref4}. 

\section{Proposed Blind Channel Estimation Method}
In this section, we discuss a block-wise transmission scheme by dividing the transmission time interval into multiple blocks and changing the phase shift matrix of RIS among blocks. In doing so, the estimation of transmit signals $\mathbf{x}_k$ and the equivalent channel $\boldsymbol{\psi}^T \mathbf{H}_k$ can be modeled as a sparse signal recovery problem. The permutation ambiguity among users is further eliminated by the identity bits embedded in the transmit signals. Moreover, the sparse cascaded channel $\mathbf{H}_k$ is estimated for further optimizations. 
\subsection{Block-wise transmission scheme}
Under the assumption that the coherence time is much longer than a codeword ($T_C\gg M$), the UE-RIS, RIS-BS channels remains unchanged during the transmission of a series of codewords. However, for accurately estimating the cascaded channel $\mathbf{H}_k$, the reflecting elements of RIS need to be reconfigured for each time slot. This means that the equivalent channel $\boldsymbol{\psi}^T \mathbf{H}_k$ is varying during the transmission time $M$ of a codeword $\mathbf{x}_k$, so that the traditional blind channel estimation model of (\ref{y(m)}) is not feasible. In order to handle the varying channel and also utilize the reconfiguration of RIS coefficients, we propose a block-wise transmission strategy. To be specific, we use a time period $T\leq T_C$ to execute the blind channel estimation and divide the time $T$ into $J$ blocks with $M$ time slots in each block, i.e., $T=MJ$. It means that different code bit vector $\mathbf{b}_k(j)$ and different codeword $\mathbf{x}_k(j)$ are transmitted in each block. Moreover, we assume that the reflecting coefficients of RIS elements differ among blocks but remain the same in one block. Then the phase shift matrix can be expressed as 
\begin{equation}\label{RISMatrix}
  \boldsymbol{\Psi}=[\boldsymbol{\psi}(1),\ldots,\boldsymbol{\psi}(J)]\in \mathbb{C}^{N_{\text{R}} \times J}
\end{equation}
 where $\boldsymbol{\psi}(j)=[\psi_1(j),\ldots,\psi_{N_R}(j)]^T$ denotes the phase shifts of all RIS elements in the $j$-th block. Hence, according to (\ref{y(m)}), we can re-express the received signals in $j$-th block $\mathbf{Y}(j)\in \mathbb{C}^{M\times N_\text{B}}$ as
  \vspace{-0.2cm}
 \begin{eqnarray}\label{Y(j)}
  \mathbf{Y}(j)
  &=&\sum_{k=1}^{K}\mathbf{Y}_k(j)\nonumber \\
  &=&[\mathbf{x}_1(j),\ldots,\mathbf{x}_K(j)] \mathbf{G}(j)  +\mathbf{N}(j)\nonumber \\
  &=&\mathbf{C}\boldsymbol{\Lambda}(j)+\mathbf{N}(j)
\end{eqnarray}
where we denotes $\mathbf{G}(j)\triangleq[\mathbf{g}_1(j),\ldots,\mathbf{g}_K(j)]^T
=[(\boldsymbol{\psi}(j)^T\mathbf{H}_1)^T,\ldots,(\boldsymbol{\psi}
(j)^T\mathbf{H}_K)^T]^T\in \mathbb{C}^{K\times{N_\text{B}}}$ as the equivalent channel from users to BS via RIS and $\boldsymbol{\Lambda}(j)=[\boldsymbol{\gamma}_1(j),\ldots,  \boldsymbol{\gamma}_K(j)]\mathbf{G}(j)$ is a row sparse matrix with $K$ non-zeros rows which respectively corresponding to $\mathbf{g}_k(j)^T,k=1,\ldots,K$. It is worth mentioning that those $K$ non-zero row indices are exactly the same with the indices of non-zero elements of $\boldsymbol{\gamma}_1(j),\ldots,\boldsymbol{\gamma}_K(j)$ in the $j$-th block. Hence we can obtain both the equivalent channel $\mathbf{G}(j)$ and the transmit signals $\mathbf{x}_k(j)$ in corresponding block by estimating matrix $\boldsymbol{\Lambda}(j)$. 

\subsection{Transmit Signal and Equivalent Channel Estimation}
According to the compressed sensing theory, the simultaneous orthogonal matching pursuit (S-OMP) algorithm \cite{ref_SOMP} can be employed to estimate the sparse matrix $\boldsymbol{\Lambda}(j)$, where the codebook $\mathbf{C}\in\mathbb{C}^{M\times N}$ is considered as the sensing matrix in each block. In this algorithm, the received signal $\mathbf{Y}(j)$ as well as the sensing matrix are used as inputs to obtain the row indices $\{\hat{n}_{p_1(j)}(j)\ldots,\hat{n}_{p_K(j)}(j)\}$ which indicate the estimation of transmit signals and the value of equivalent channel $\{\hat{\mathbf{g}}_{p_1(j)}(j),\ldots,\hat{\mathbf{g}}_{p_K(j)}(j)\}$, where $p_k(j)$ denotes the $k$-th output from the support during S-OMP process in the $j$-th block. It is intuitive that a larger $M$, i.e., a longer codeword, results in a better recovery accuracy in this compressed sensing algorithm. 

We note that, since the received signal is a superposition of signals arrived at BS from all $K$ users, the recovery of transmit signals and channels have a permutation ambiguity among users, which has been widely studied as an inherent property in blind estimation problems\cite{BSD_mMIMO}\cite{UMA_BCE}. This means that reordering of subscripts $\{p_1(j),\ldots,p_K(j)\}$ is required additionally. In order to eliminate the permutation ambiguity and accurately identify the users, identity bits (ID bits) need to be embedded into the code bit vectors. For $K$ users system, $M_K=\lceil \log_{2}{K} \rceil$ bits are required to identify each user. For instance, we consider a 4-users system where $M_K=2$ identity bits are required. We can choose from a set of 2-bits codes $\mathcal{B}=\{00,01,10,11\}$ as the first 2 bits of each user's code bit vector $\mathbf{b}_k(j)$. 
Then the value of $n_k$ is restricted in a specific range for each user. Hence, the receiver is capable to permute $\{\hat{n}_{p_1(j)}(j)\ldots,\hat{n}_{p_K(j)}(j)\}$ into correct order.
After we obtain the correct permutation, which can be denoted as $\{\hat{n}_1(j)\ldots,\hat{n}_K(j)\}$, we can express the transmit signal as
\begin{equation}\label{x_hat}
  \hat{\mathbf{x}}_k(j)=\mathbf{C}(:,\hat{n}_k(j))
\end{equation}
Then, according to the mapping relationship from $\{p_1(j),\ldots,p_K(j)\}$ to $\{1,\ldots,K\}$, the equivalent channel $\{\hat{\mathbf{g}}_{p_1(j)}(j),\ldots,\hat{\mathbf{g}}_{p_K(j)}(j)\}$, which are obtained as the non-zero rows of $\boldsymbol{\Lambda}(j)$ can be permuted in the same way, which compose the equivalent channel matrix 
$\hat{\mathbf{G}}(j)=[\hat{\mathbf{g}}_1(j),\ldots,\hat{\mathbf{g}}_K(j)]^T$. 

\subsection{Cascaded Sparse Channel Estimation}
After we obtain the transmit signal $\hat{\mathbf{x}}_k(j)$ and equivalent channel $\hat{\mathbf{g}_k}(j)$, it is still not enough for further system optimization \cite{RIS_BF}. As denoted earlier in (\ref{H_kSparse}), the sparse matrix $\mathbf{D}_k$, containing the path gains and angle information of cascaded channel $\mathbf{H}_k$, is required. Combining (\ref{H_kSparse}) and (\ref{Y(j)}), the equivalent channel for user $k$ in $j$-th block $\mathbf{g}_k(j)\in \mathbb{C}^{N_B\times 1}$ can be denoted as 
\begin{equation}\label{g_K(j)}
  \mathbf{g}_k(j)^T=\boldsymbol{\psi}(j)^T\mathbf{H}_k
  =\boldsymbol{\psi}(j)^T\mathbf{F}_{\text{R}}
  \mathbf{D}_k\mathbf{F}_{\text{B}}^H
\end{equation}
Using the property of Kronecker product, (\ref{g_K(j)}) can be rewritten as
\begin{equation}\label{vec(g_k(j))}
  \mathbf{g}_k(j)=\left(\mathbf{F}_{\text{B}}^*\otimes
  (\boldsymbol{\psi}(j)^T\mathbf{F}_{\text{R}})\right)\text{vec}(\mathbf{D}_k)
\end{equation}
where $\text{vec}(\cdot)$ denotes the vectorization of a matrix. To estimate the cascaded channel, we collect all $J$ blocks of user-$k$'s channel with varying phase shifts of RIS, which is shown as
\begin{equation}\label{OMP_formulation}
  \mathbf{g}_k=\left[
  \begin{array}{c}
    \mathbf{F}_{\text{B}}^*\otimes \left(\boldsymbol{\psi}(1)^T\mathbf{F}_{\text{R}}\right) \\
    \vdots \\
    \mathbf{F}_{\text{B}}^*\otimes
  \left(\boldsymbol{\psi}(J)^T\mathbf{F}_{\text{R}}\right) 
  \end{array}
  \right]\mathbf{d}_k
\end{equation}
where $\mathbf{g}_k=[\mathbf{g}_k(1)^T,\ldots,\mathbf{g}_k(J)^T]^T\in \mathbb{C}^{JN_{\text{B}}\times1}$, $\mathbf{d}_k=\text{vec}(\mathbf{D}_k)$ is the channel to be estimated. Here the sensing matrix is expressed as 
\begin{equation}\label{Q}
  \mathbf{Q}=\left[
  \begin{array}{c}
    \mathbf{F}_{\text{B}}^*\otimes \left(\boldsymbol{\psi}(1)^T\mathbf{F}_{\text{R}}\right) \\
    \vdots \\
    \mathbf{F}_{\text{B}}^*\otimes
  \left(\boldsymbol{\psi}(J)^T\mathbf{F}_{\text{R}}\right) 
  \end{array}
  \right]\in \mathbb{C}^{JN_{\text{B}}\times{G_{\text{B}}G_{\text{R}}}}
\end{equation}
Therefore, the sparse cascaded channel estimation problem has been converted into a sparse signal recovery problem which is easy to be solved by employing compressed sensing algorithms. In our study, we use the orthogonal matching pursuit (OMP) algorithm to recover the sparse signal $\mathbf{d}_k$. 

Furthermore, according to the well-known compressed sensing theory and related works applied to communication systems, the sensing matrix can achieve a better recovery accuracy when it is properly designed. We utilize a manifold optimization to minimize the column mutual coherence of the sensing matrix $\mathbf{Q}$ to generate an approximately orthogonal dictionary. The column mutual coherence of $\mathbf{Q}$ is defined as 
\begin{equation}\label{mutual_coherence}
    \mu(\mathbf{Q})\triangleq \mathop{\max}_{1\leq m \textless n \leq G_\text{B}G_\text{R}}\frac{\mathbf{q}_m^H\mathbf{q}_n}{\Vert\mathbf{q}_m\Vert_2\Vert\mathbf{q}_n\Vert_2}
\end{equation}
where $\mathbf{q}_m$ and $\mathbf{q}_n$ denote the $m$-th and $n$-th column of $\mathbf{Q}$, respectively. For better illustration, we introduce an auxiliary variable $\tilde{\mathbf{Q}}\triangleq\left[\mathbf{F}_\text{B}^*\otimes(\boldsymbol{\Psi}^T\mathbf{F}_\text{R})\right]$. Based on the properties of the Kronecker product, $\tilde{\mathbf{Q}}$ is actually a row reordered form of $\mathbf{Q}$. Intuitively, adjustment on row permutation has no influence on the column mutual coherence so $\mu(\mathbf{Q})=\mu(\tilde{\mathbf{Q}})$. Furthermore, due to the Kronecker product structure, the two part of $\mathbf{F}_\text{B}^*$ and $\boldsymbol{\Psi}^T\mathbf{F}_\text{R}$ are only related to the BS side and RIS-user side, respectively, so the mutual coherence in (\ref{mutual_coherence}) can be further decoupled as $\mu(\tilde{\mathbf{Q}})=\mathop{\max}\{\mu({\mathbf{F}_\text{B}^*}),\mu(\boldsymbol{\Psi}^T\mathbf{F}_\text{R})\}$\cite{R4}.

As an array response matrix, $\mathbf{F}_\text{B}^*$ has already achieved a good column mutual coherence property. So we focus on the RIS-user part component $\boldsymbol{\Psi}^T\mathbf{F}_\text{R}$ and the optimization problem can be further stated as\cite{R6}
\begin{align}\label{R5}
       \mathop{\min}_{\boldsymbol{\Psi}}\quad&\Vert \mathbf{F}_\text{R}^H\boldsymbol{\Psi}^*\boldsymbol{\Psi}^T\mathbf{F}_\text{R}-\xi \mathbf{I}_{G_\text{R}}\Vert_F^2\nonumber \\
    \text{s.t.}\quad&\vert\boldsymbol{\Psi^T}\vert=\mathbf{1}_{J\times N_\text{R}}
\end{align}
where $\mathbf{F}_\text{R}^H\boldsymbol{\Psi}^*\boldsymbol{\Psi}^T\mathbf{F}_\text{R}$ is a symmetric and positive semi-definite Gram matrix \cite{R5} and $\xi$ is a normalization parameter which is introduced for reducing the impact of the elements on the main diagonal of the Gram matrix. Hence, by defining the main diagonal of $\mathbf{F}_\text{R}^H\boldsymbol{\Psi}^*\boldsymbol{\Psi}^T\mathbf{F}_\text{R}$ as a vector $\boldsymbol{\eta}\in\mathbb{C}^{G_\text{R}\times 1}$, a new optimization problem for $\xi$ is introduced as
\begin{equation}\label{R6}
    \mathop{\min}_{\xi}\Vert\boldsymbol{\eta}-\xi\mathbf{1}_{G_{\text{R}}\times 1}\Vert_F^2
\end{equation}
which is constrained according to the RIS pattern matrix $\boldsymbol{\Psi}$. It is obvious that an optimal solution for (\ref{R6}) is $\xi^*=\frac{\text{tr}(\mathbf{F}_\text{R}^H\boldsymbol{\Psi}^*\boldsymbol{\Psi}^T\mathbf{F}_\text{R})}{G_{\text{R}}}$ obtained through a gradient calculation. Then, considering the design of non-diagonal elements of $\mathbf{F}_\text{R}^H\boldsymbol{\Psi}^*\boldsymbol{\Psi}^T\mathbf{F}_\text{R}$ in (\ref{R5}). Noticing that the constraint $\vert\boldsymbol{\Psi^T}\vert=\mathbf{1}_{J\times N_\text{R}}$ in (\ref{R5}) could be seen as a complex circle domain (or Riemannian manifold), so it is capable to be solved by using the manifold optimization theory. A detailed approach is elaborated in \cite{R7} which is feasible in our paper to design the RIS pattern. More detailed information about manifold optimization can be found in \cite{R8} and we omit it due to the length limitation of the manuscript. The overall procedure is summarized in Algorithm \ref{alg.1}. 
\begin{algorithm}[]
\caption{Proposed Blind Channel Estimation Algorithm.}\label{alg.1}
\begin{algorithmic}[1]
\REQUIRE
$\mathbf{Y}$, $\mathbf{C}$,$\boldsymbol{\Psi}$, $\mathbf{F}_{\text{B}}$ $\mathbf{F}_{\text{R}}$, $K$, $L_{\text{RB}}$, $L_{\text{RU,k}}$
\ENSURE
Cascaded channel $\hat{\mathbf{H}}_k=\mathbf{F}_{\text{R}}
  \hat{\mathbf{D}}_k\mathbf{F}_{\text{B}}^H$, transmit signals $\hat{\mathbf{x}}_k(j)=\mathbf{C}(:,\hat{n}_k(j))$, $k=1,\ldots,K,j=1,\ldots,J$
\FOR {$j=1,\ldots,J$}
\STATE Estimating $\boldsymbol{\Lambda}(j)$ by S-OMP algorithm\cite{UMA_BCE}. Initialization: $\mathbf{R}=\mathbf{Y}(j)^T$, $\mathcal{I}=\emptyset$
\FOR {$t=1,\ldots,N$}
\STATE $n^*=\mathop{\argmax}_{n\in 1,\ldots,N}\frac{\Vert\mathbf{R}^H\mathbf{C}(:,n)\Vert_2}{\Vert\mathbf{C}(:,n)\Vert_2}$
\STATE $\mathcal{I}\leftarrow\mathcal{I}\cup{n^*}$
\STATE $\boldsymbol{\Sigma}=\mathbf{C}(:,\mathcal{I})^{\dagger}\mathbf{Y}(j)^T$
\STATE $\mathbf{R}=\mathbf{Y}(j)^T-\mathbf{C}(:,\mathcal{I})^{\dagger}\boldsymbol{\Sigma}$
\ENDFOR
\STATE Find $K$ rows of $\boldsymbol{\Sigma}$ with the largest $l_2$ norms $\{\hat{\mathbf{g}}_{p_1(j)}(j),\ldots,\hat{\mathbf{g}}_{p_K(j)}(j)\}$ and the corresponding indices $\{\hat{n}_{p_1(j)}(j)\ldots,\hat{n}_{p_K(j)}(j)\}$
\STATE Eliminate the permutation ambiguity to obtain $\{\hat{n}_1(j)\ldots,\hat{n}_K(j)\}$ and $\{\hat{\mathbf{g}}_1(j),\ldots,\hat{\mathbf{g}}_K(j)\}^T$
\ENDFOR
\FOR {$k=1,\ldots,K$}
\STATE Obtain $\mathbf{g}_k$ by collect $\mathbf{g}_k(j)$, and calculate $\mathbf{Q}$ by (16)
\STATE Obtain $\hat{\mathbf{D}}_k$ by OMP algorithm\cite{ref4}. Initialization: $\mathbf{r}_0=\mathbf{g}_k$, $\boldsymbol{\Lambda}_0=\emptyset$
\FOR {$l=1,L_{\text{RB}}L_{\text{RU},k}$}
\STATE $\lambda_l=\mathop{\argmax}_{j\notin\Lambda_{l-1}}\vert\mathbf{Q}(:,j)^T\mathbf{r}_{l-1}\vert$
\STATE $\Lambda_{l}=\Lambda_{l-1}\cup\lambda_l$
\STATE $\mathbf{d}_k(l)=\mathop{\argmin}_{\mathbf{d}_k}\Vert\mathbf{Q}(:,\Lambda_{l})\mathbf{d}_k-\mathbf{g}_k\Vert$
\STATE $\mathbf{r}_l\leftarrow\mathbf{g}_k-\mathbf{Q}\mathbf{d}_k(l)$
\ENDFOR 
\STATE Calculate the cascade channel $\hat{\mathbf{H}}_k$ for each user 
\ENDFOR 
\end{algorithmic}
\end{algorithm}

\subsection{Complexity and Overhead Analysis}
Algorithm 1 is mainly composed by three parts, namely {\it (i)} S-OMP algorithm for transmit signal and equivalent channel recovery, {\it (ii)} sorting for permutation ambiguity elimination, {\it (iii)} OMP algorithm for cascade channel estimation. During the S-OMP algorithm, the computational complexity is dominated by the pseudo-inverse operations [9], which is $\mathcal{O}(M^3)$ in $N$ times iterations in $J$ blocks in total. The computational complexity when dealing with the permutation ambiguity in step 3 is dominated by sorting the estimated column indices into correct order, which is no more than $\mathcal{O}(K^2)$ in each block. And the OMP algorithm has the computational complexity dominated by the calculation of inner product in step 9 \cite{ref4} which is $\mathcal{O}(G_{\text{R}}G_{\text{B}}N_{\text{B}}J)$ and the operation of pseudo-inverse in step 11 which is $\mathcal{O}(N_{\text{B}}JL_k^2)$ in $L_k$ times iterations for each of the $K$ users, where $L_k=L_{\text{RU},k}L_{\text{RB}}$ denotes the sparsity of channel for each user. Hence the overall computational complexity is approximately $\mathcal{O}(NM^{3}J+K^{2}J+(G_{\text{R}}G_{\text{B}}+L^2)N_{\text{B}}JKL_k)$. 
\begin{table*}[bt]
  \centering
  \caption{Complexity, overhead, and data rate Comparison}  \label{Complexity}
 \begin{tabular}{|c|c|c|c|c|}
    \hline
    Channel Estimation Method & Category & Computational Complexity & Pilot/ID overhead & Data rate (bps) \\
    \hline
    Pilot-based Method in [4] & pilot-based & $\mathcal{O}(G_{\text{R}}G_{\text{B}}+L^2)N_{\text{B}}KTL)$ &$\mathcal{O}(\log{G_\text{R}G_\text{B}}\sum_k{L_k})$ & $\geq (1-\frac{T_p}{T})$ \\ \hline 
    TALS Method in [10] & semi-blind & $\mathcal{O}(N_{\text{R}}^2MJ(1+N_{\text{B}}^2K)+N_{\text{B}}JK(N_{\text{R}}M+N_{\text{B}}))$ & $JK$ & $\frac{4\times K\times (M_b-1)}{M\times J}$\\ \hline
    Proposed Method & blind & $\mathcal{O}(NM^{3}J+JK^{2}+(G_{\text{R}}G_{\text{B}}+L^2)N_{\text{B}}JKL)$ & $J\log_{2}{K}$ & $\frac{K\times (M_b-\log_2K)}{M}$\\
    \hline
  \end{tabular}
\end{table*}

For comparison, Table I summarize the computational complexity and training overhead of our proposed method and the existing channel estimation methods, i.e., the pilot-based method\cite{ref4} and the tensor-based trilinear alternating least square (TALS) semi-blind method\cite{ref5}. Specifically, the conventional pilot-based method uses orthogonal pilots for all potential users and the estimation algorithm is also based on compressed sensing. The TALS method in \cite{ref5} utilize a block division scheme to construct a tensor model for the received signals. Although it is different from our proposed method, we also use the parameter $J$ to denote the number of blocks and $M$ to denote the length of each block for ease of expression. Our proposed method exhibits an advantage on computational complexity, especially when the number of RIS elements grows larger. 

Moreover, we would like to discuss about the overhead and the estimation performance. As introduced in Section III-B, $M_K=\lceil \log_{2}{K}\rceil$ bits are required to identify each user in the code bit vector. The total overhead is $J\times{M_K}$, which is increased logarithmically with the number of users. As comparison, the traditional pilot-based method requires at least $\log{G_\text{R}G_\text{B}}\sum_kL_k$ pilots to estimate the cascade channel which is increased linearly with the number of users. The overhead for handling the scaling and permutation ambiguities for TALS method in [10] is $JK$. In addition, the transmission data rate of our proposed method out performs the semi-blind method in [10]. That is reasonable since we transmit different data sequences among different blocks while the referenced one repeatedly transmit the same data stream among all $J$ blocks. Since the exact modulation approach of pilot-based method in [4] is unmentioned, the data rate $1-\frac{T_p}{T}$ is under the assumption of BPSK modulation. 
Although its data rate is capable to increase when utilizing more effective encoding approaches with high modulation order, the UMA scenario with a large amount of potential users makes the required pilot length $T_p$ extremely long\cite{UMA_BCE}. It will lead to reduced data rate per channel use.

\section{Simulation Results}
In this section, we present our simulation results to evaluate the performance of proposed blind channel estimation method. In this work, we consider an RIS-assisted multi-user mmWave system, where the number of RIS elements is set to $N_{\text{R}}=32$ and the BS is equipped with $N_{\text{B}}=4$ antennas. Without loss of generality, we assume that each user to RIS channel has the same paths number, which can be expressed as $L_{\text{RU},k}=L_{\text{RU}}, \forall k=1,\ldots,K$. Then the number of paths for $\mathbf{H}_\text{RB}$ and $\mathbf{h}_{\text{RU},k},k=1,\ldots,K$ are set to $L_{\text{RB}}=L_{\text{RU}}=2$, and the corresponding path gains are respectively set to $\rho_l\sim\mathcal{CN}(0,\frac{1}{L_{\text{RB}}})$,
$\varrho_{l,k}\sim\mathcal{CN}(0,\frac{1}{L_{\text{RU}}})$. The AoAs/AoDs are uniformly chosen from the discretized grid\footnote{like many other works\cite{ref4}\cite{ref7}, we assume that the actual angles lie on the discretized grids for simplicity. Considering the practical off-grid problem which means the angles not necessarily lie on the grids\cite{off-grid}, the proposed algorithm is capable to be refined by adding a gradient descent procedure after the cascade channel estimation.} and the number of grids at the BS side and the RIS side are set to $G_{\text{B}}=16$ and $G_{\text{R}}=64$, respectively. The codebook $\mathbf{C}$ is a randomly generated complex Gaussian matrix consisting of i.i.d. $\mathcal{CN}(0,1)$ elements. The length of code bit vector of each user in each block is $M_b=8$, and thus there are $N=2^{8}=256$ codewords in the codebook. 

In order to evaluate the performance of the proposed blind channel estimation scheme, we use NMSE defined as $\mathbb{E}\left[\|\mathbf{H}_k-\hat{\mathbf{H}}_k\|^2_F/\|\mathbf{H}_k\|^2_F\right]$. During the process of transmit signal recovery, the signal $\mathbf{x}_k(j)$ and code vector $\mathbf{b}_k(j)$ can be accurately found as long as the row index $n_k(j)$ is correctly estimated. Noting that once an error occurs in the ID bits, the data would be transmitted to the wrong user and the entire bit sequence would become useless. Hence, the weight of the ID bits is $M_b$ times greater than that of data bits in terms of bit error rate (BER). 

Fig. 1 shows the weighted BER for the transmit signal recovery against signal-to-noise ratio (SNR) under different condition of the length of codebook $M$, which also represents the row number of the sensing matrix for S-OMP algorithm. Simulation results show that our blind channel estimation scheme exhibits excellent signal recovery performance when the codebook length achieves a certain range. 
\begin{figure}[t]
\setlength{\abovecaptionskip}{-0.2cm}
\setlength{\belowcaptionskip}{-0.8cm}
  \centering
  \includegraphics[width=3in]{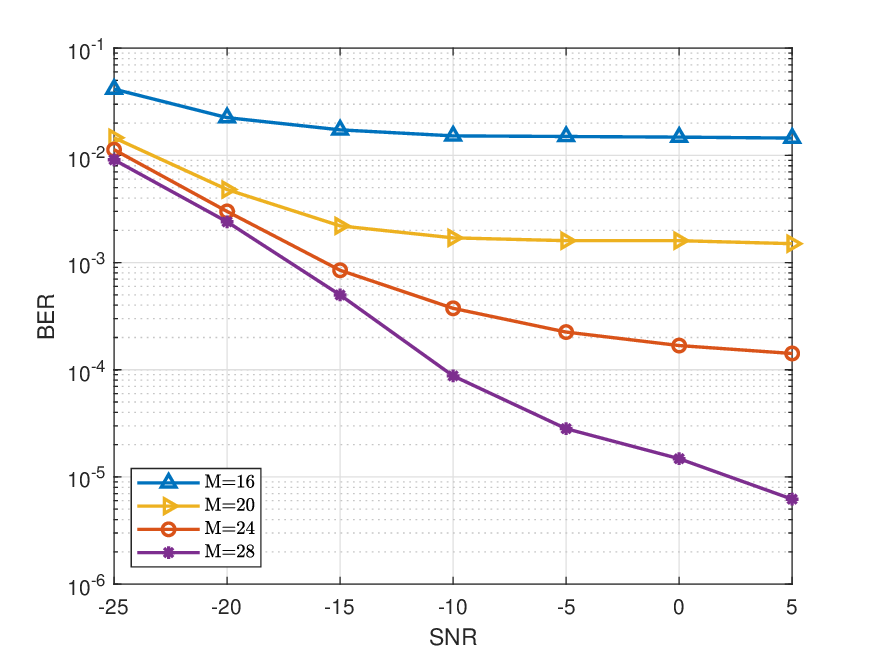}
  \caption{Transmit signal recovery performance against SNR with $J=60,K=4$. }\label{BER_SNR}
\end{figure}
\begin{figure}[t]
\setlength{\abovecaptionskip}{-0.2cm}
\setlength{\belowcaptionskip}{-0.8cm}
  \centering
  \includegraphics[width=3in]{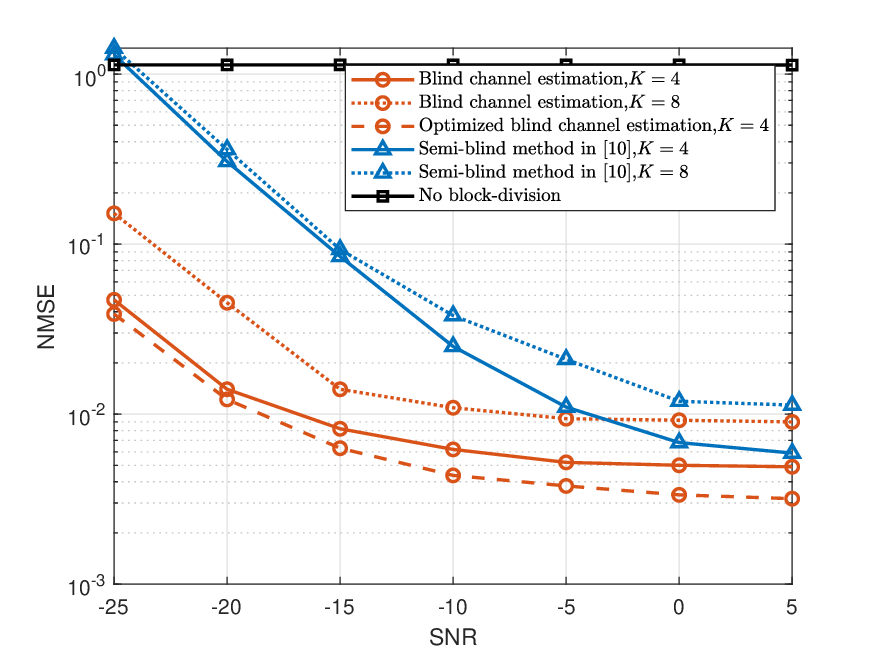}
  \caption{Cascaded channel estimation performance against SNR. }\label{NMSE_SNR}
\end{figure}

We plot the NMSE performance in Fig. 2, where the red lines represent the proposed blind channel estimation and the blue lines represent the semi-blind method in [10] as comparisons. Different line styles indicate different setup of parameters. A longer codeword (a larger $M$) intuitively leads to a better estimation accuracy but a slightly decreasing transmission data rate. Hence, the value of $M$ should be selected moderately.
To be specific, the red solid line denotes a 4-user system where the length of codebook is $M=28$ with $J=30$ transmission blocks. Since $M_{K}=\lceil \log_{2}{4}\rceil=2$ ID bits are involved in each code bit vector, $\frac{M_k}{M_b}=25\%$ time resource is occupied as the overhead of estimation. Similarly, the red dotted line present the estimation performance of a 8-user system. The corresponding parameters of the semi-blind method in [10], which is shown as the blue lines, are selected to ensure the pilot takes the same ratio of transmission symbols as well as the same transmission time slots being occupied.
Intuitively, the performance of the proposed blind channel estimation slightly decreases as the number of users grows, because the codebook size $M$ and block number $J$ are fixed. We also observe that our proposed method outperform the conventional semi-blind methods. Besides, it shows that the optimization of RIS phase shift matrix further improve the performance of channel estimation.

For comparison, the black dotted line shows the scenario without block-division, which means $M=T,J=1$ and the phase shift matrix of RIS remain fixed among all the time slots. It could be considered as a conventional blind channel estimation scheme, where the equivalent channel remains unchanged during the coherence time. However, it failed to estimate the sparse cascaded channel $\mathbf{D}_k$ from the equivalent channel $\mathbf{g}_k$ because the fixed RIS actually act no help on the estimation of cascade channel.

\section{Conclusions}
This paper considered the blind channel estimation for the uplink of an RIS-assisted multiuser mmWave communication system. We proposed a method to estimate the transmit signals generated from a common codebook shared among users and the cascade channel at the same time. By dividing the transmission time into multiple blocks, our blind channel estimation approach can accurately recover the transmit signals for each user in different blocks and estimate the cascaded channels by combining multiple blocks, among which RIS elements are reconfigured. Simulation results showed that our method achieves accurate blind channel estimation and signal recovery, and also outperform the traditional semi-blind methods with a lower overhead.

\newpage

\vfill

\end{document}